# Bragg-type soliton mirror

Yaroslav V. Kartashov,[1] Victor A. Vysloukh,[2] and Lluis Torner[1]

[1]*ICFO-Institut de Ciencies Fotoniques, and Universitat Politecnica de Catalunya, Mediterranean Technology Park, 08860 Castelldefels (Barcelona), Spain*
[2]*Departamento de Física y Matemáticas, Universidad de las Americas – Puebla, Puebla 72820, Mexico*
*Yaroslav.Kartashov@icfo.es*

**Abstract:** We study soliton reflection/transmission at the interface between uniform medium and the optical lattice with focusing Kerr nonlinearity. We reveal that such interfaces afford rich new opportunities for controlling the reflection and transmission coefficients and nonlinear Snell law, the key control parameters being the spatial frequency and depth of the lattice.

## References and links

The interface between two different nonlinear media has been the subject of study in nonlinear optics during the last three decades. Nonlinearity substantially affects the classical Snell law of refraction of planar waves as well as Fresnel's reflection formulas and can lead to optical bistability and hoping between regimes of total internal reflection and complete transmission [1,2]. The influence of nonlinear interface on light refraction was studied for focused laser beams both theoretically [3-6] and experimentally [7-10], in cubic and in quadratic nonlinear media. Laser beams can excite localized surface waves at a single interface if one of the materials forming the interface possesses nonlocal nonlinear response [11]. Surface waves can also be formed at the interface of linear dielectric and periodic layered media [12].

Beam refraction at nonlinear interfaces can potentially be used for beam steering, routing, and switching. Photorefractive materials and nonlinear semiconductors (such as AlGaAs) offer an excellent opportunity for experimentation at moderate power levels. The important fact is that photorefractive materials can be used for optical induction of photonic lattices [13-16], while nonlinear semiconductors allows the fabrication of the arrays of weakly coupled waveguides [17-19]. Such guiding structures can support lattice or discrete solitons whose properties depart considerably from properties of solitons in the uniform media [13-22]. Thus, tunable beam refraction in infinite optically induced tilted lattice was recently demonstrated experimentally in Ref. [23].

Light refraction on nonlinear interfaces is far from being trivial even for the simplest case of uniform materials, and it has not yet been explored in a more challenging case of the interface formed by nonlinear uniform media and periodic guiding structures. Such structures are practically interesting since they introduce several experimentally controllable parameters, namely the features of the lattice. From a fundamental point of view, they are also intriguing because the eigenfunctions that correspond to the nonlinear uniform medium are those of the Zakharov-Shabat inverse scattering problem (i.e., continuous Fourier spectrum plus discrete eigenvalues linked with the single or multiple solitons), while the eigenfunctions associated in the linear limit with photonic lattices are Floquet-Bloch modes. In such a hybrid situation the energy flow of soliton falling upon the interface is redistributed between two groups of eigen-functions mentioned above depending on the interface properties, i.e., the lattice depth and frequency. Understanding and controlling this redistribution is in the focus of this paper.

We address refraction/reflection of tilted laser beams at the interface of a uniform Kerr-type nonlinear medium and a nonlinear medium with fabricated or optically induced harmonic transverse modulation of refractive index. Since the ability of laser beams to penetrate and to propagate in the periodic materials depends strongly on the depth and frequency of refractive index modulation, it is natural to expect that refraction scenario at such interface departs crucially from refraction at interface of two uniform media and allow for new regimes of light switching and routing.

For the sake of generality we consider propagation of tilted slit laser beam at an interface of uniform and periodic materials with Kerr-type cubic focusing nonlinearity described by the nonlinear Schrödinger equation for the dimensionless complex amplitude of light field $q$:

$$i\frac{\partial q}{\partial \xi} = -\frac{1}{2}\frac{\partial^2 q}{\partial \eta^2} - q|q|^2 - pR(\eta)q. \qquad (1)$$

Here $q(\eta,\xi) = (L_{\text{dif}}/L_{\text{nl}})^{1/2} A(\eta,\xi) I_0^{-1/2}$, $A(\eta,\xi)$ is the slowly varying envelope of the light field, $I_0$ is the input intensity, $\eta = x/x_0$, $x_0$ is the width of input beam, $\xi = z/L_{\text{dif}}$ is the normalized propagation distance, $L_{\text{dif}} = n_0 \omega x_0^2 / c$ is the diffraction length, $\omega$ is the carrying frequency, $L_{\text{nl}} = 2c/\omega n_2 I_0$ is the nonlinear length, $p = L_{\text{dif}}/L_{\text{ref}}$ is the guiding parameter of the lattice, $L_{\text{ref}} = c/(\delta n \omega)$ is the linear refraction length, $\delta n$ is the refractive index modulation depth. The function $R(\eta)$ stands for the refractive index profile and is given by $R(\eta) = 1 - \cos(\Omega\eta)$ at $\eta \geq 0$ and $R(\eta) = 0$ at $\eta < 0$, where $\Omega$ is the lattice frequency. We assume that the depth of linear refractive index modulation is small compared with the unperturbed index $n_0$ and is of the order of the nonlinear correction due to Kerr effect. This regime is most interesting because in too deep lattices one recovers the quasi-linear regime well studied in the context of Bragg gratings and multi-layered mirrors, while shallow lattices cannot substantially modify the refraction law for intense laser beams. Such specific refractive index landscapes $R(\eta)$ can be either technologically fabricated in semiconductors that were used for creation of discrete waveguide arrays [17-19], or they can be induced optically by the interference pattern formed in the photorefractive crystal, provided that the refractive index of one part of the crystal is modified, for example, by the ion implantation. In the latter case vectorial interactions can be used to guide soliton beam in optically induced lattices [13-16], while lattice parameters can be tuned by the interference pattern. Note, that Eq. (1) conserves the energy flow

$$U = \int_{-\infty}^{\infty} |q|^2 \, d\eta. \qquad (2)$$

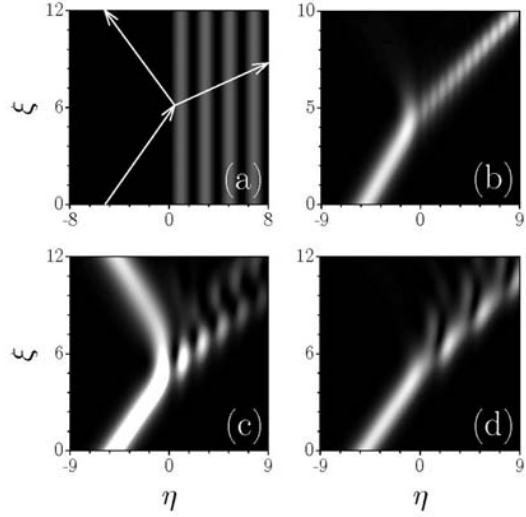

Fig. 1. (a) Schematic representation of the geometry of soliton refraction at the boundary between the lattice and uniform medium. Soliton propagation dynamics in illustrative cases, at $\Omega = 8$ (b), 3.5 (c), and 2 (d). In all cases $\alpha_{\text{in}} = 1$, $p = 1$, $\chi = 1$.

To elucidate the refraction scenarios at such interface we integrated Eq. (1) by the beam propagation method with input conditions $q(\eta,\xi=0)=\chi\,\mathrm{sech}[\chi(\eta+\eta_0)]\exp[i\alpha_{\mathrm{in}}(\eta+\eta_0)]$, corresponding to a single soliton with form-factor $\chi$, launched into the uniform nonlinear medium at distance $\eta_0 \gg \chi^{-1}$ from the interface. The positive incident angle $\alpha_{\mathrm{in}}$ (that we calculate from the interface plane) corresponds to soliton motion toward the interface (see Fig. 1(a)). We consider only the case $p > 0$ that corresponds to external beam reflection, when the mean value of linear refractive index inside the lattice exceeds the refractive index $n_0$ of the uniform medium. For typical soliton widths and modulation periods of the order of tenths

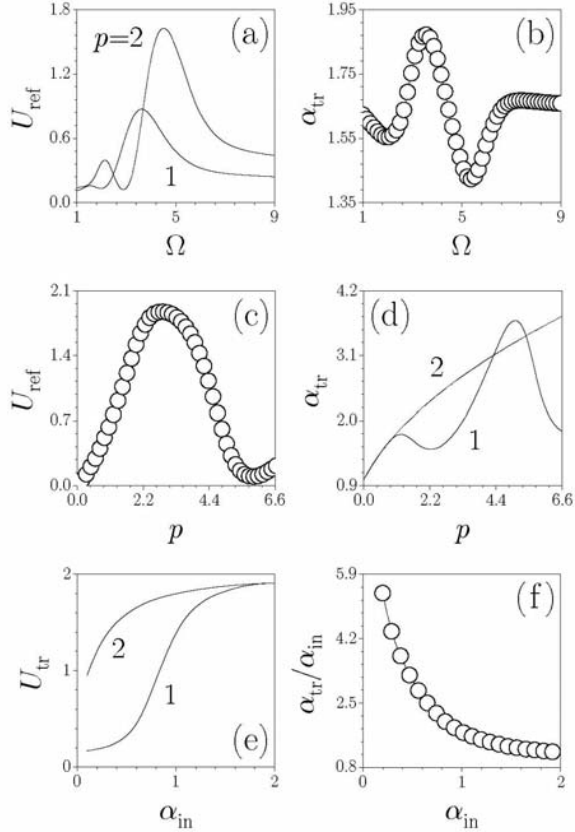

Fig. 2. (a) Reflected energy vs lattice frequency at $\alpha_{\mathrm{in}}=1$. (b) Propagation angle of transmitted soliton vs lattice frequency at $\alpha_{\mathrm{in}}=1$, $p=1$. (c) Reflected energy vs lattice depth at $\alpha_{\mathrm{in}}=1$, $\Omega=4.2$. (d) Propagation angle of transmitted soliton vs lattice depth at $\alpha_{\mathrm{in}}=1$, $\Omega=4.2$. Curve 1 – numerical results, curve 2 – analytical effective index approximation. (e) Energy of transmitted soliton vs incident angle at $p=1$, $\Omega=3$. Curve 1 – numerical results, curve 2 – analytical effective index approximation. (f) Ratio of propagation angles of transmitted and incident solitons versus incident angle at $p=1$, $\Omega=3$. In all cases $\chi=1$.

microns the diffraction length is of the order of a few millimeters and propagation angles in radians can be obtained from the normalized ones with multiplication by the scaling factor $x_0/L_{\mathrm{dif}} \ll 1$. Since one of the physically meaningful factors that determine the propagation regime of laser beam in periodic structure is not the width of beam itself but rather the ratio of the beam width to lattice period (or ratio of lattice frequency to form-factor) [21], in numerical simulations we fixed the input soliton width (that also determines its energy flow) by setting

$\chi = 1$ and we varied the lattice depth and frequency. This allows one to recover all possible refraction scenarios at the interface between uniform medium and the lattice. The lattice depth and frequency can be tuned by adjusting the propagation angles and intensities of interfering plane waves, which induce the lattice. We assume that the lattice depth is of the order of the nonlinear contribution to the refractive index, thus we consider the range $p \in [0,10]$.

We found that the beam reflection/refraction scenario at the interface depends crucially on the lattice frequency (see Figs. 1(b)-1(d)). In the high-frequency lattice ($\Omega \gg \chi$) soliton is almost unaffected by periodic character of refractive index modulation since it covers many lattice periods and feels only mean refractive index of the lattice or, in other words, the soliton spatial spectrum does not overlap with the Bragg-reflection band. Under such conditions sech-type spatial soliton is almost totally transmitted into the lattice (Fig. 1(b)), provided that the transmission angle is less than the Bragg reflection angle $\alpha_{\text{tr}} < \Omega/2$. Notice, that light beam is transmitted with high soliton content (Fig. 1(b)), which means that almost the whole energy associated to the Zakharov-Shabat spectrum converts into soliton moving across the lattice. In this particular case, a simple analytical effective index approximation can be used to derive an estimate for the transmission angle $\alpha_{\text{tr}}$. Introducing the integral beam center as

$$\eta_{\text{c}}(\xi) = U^{-1} \int_{-\infty}^{\infty} |q|^2 \eta d\eta, \tag{3}$$

one gets the following evolution equation:

$$\frac{d^2\eta_{\text{c}}}{d\xi^2} = \frac{p}{U} \int_{-\infty}^{\infty} \frac{dR}{d\eta} |q|^2 d\eta. \tag{4}$$

When $\Omega \gg \chi$, the oscillating profile of the refractive index can be approximated by the averaged one $R(\eta) = [1 + \tanh(\Omega\eta)]/2$. At $\Omega \to \infty$ Eq. (4) transforms into equation

$$\frac{d^2\eta_{\text{c}}}{d\xi^2} = \frac{p\chi}{2\cosh^2(\chi\eta_{\text{c}})}, \tag{5}$$

whose integration with the initial conditions $\eta_{\text{c},\xi=0} = \eta_0$ and $(d\eta_{\text{c}}/d\xi)_{\xi=0} = \alpha_{\text{in}}$ yields the final estimate for soliton transmission angle $\alpha_{\text{tr}} = d\eta_{\text{c}}/d\xi = (\alpha_{\text{in}}^2 + 2p)^{1/2}$.

When lattice frequency $\Omega$ decreases, the condition of Bragg reflection $\alpha_{\text{tr}} = \Omega/2$ for transmitted beam can be met. Around this Bragg frequency the reflection coefficient growths dramatically, and upon interaction with interface soliton bounces back into uniform nonlinear media (Fig. 1(c)), so that almost the whole energy remains with the Zakharov-Shabat discrete eigenmode. For a lattice with a low frequency belonging to the Bragg-transmission band the beam reflection is suppressed (Fig. 1(d)). Notice that in this latter case radiative losses for solitons moving across the lattice are important, because when soliton crosses the successive waveguides it leaves there some small fraction of energy due to the overlap of low-frequency wing of the soliton spatial spectrum (which is centered at $\alpha_{\text{tr}}$) with spatial spectrum of guided mode that is centered at zero spatial frequency. This process is clearly visible in Fig. 1(d). The excitation of Floquet-Bloch modes is also possible, as it was demonstrated experimentally in Ref. [17], in a different regime than the one we explore here.

The key results of this paper are summarized in Fig. 2, which illustrates broad prospects for soliton reflection and refraction control. Figure 2(a) shows the key feature of such Bragg-type soliton mirror: the dependence of the reflected energy flow $U_{\text{ref}}$ (that was calculated at the distance $\xi = 2\eta_0/\alpha_{\mathrm{in}}$) on the lattice frequency $\Omega$. Let the reflected $U_{\mathrm{ref}}$ and transmitted $U_{\mathrm{tr}}$ energies

$$U_{\mathrm{ref}} = \int_{-\infty}^{0} |q|^2 d\eta \text{ and } U_{\mathrm{tr}} = \int_{0}^{\infty} |q|^2 d\eta. \tag{6}$$

Large enhancement of the reflection efficiency at $\Omega \approx \Omega_{\mathrm{B}}$ due to the distributed Bragg-type scattering and formation of well-defined reflection band should be pointed out. Notice, that in the high-frequency limit $(\Omega \gg \chi)$ energy reflection is determined primarily by the step of the averaged refractive index profile. The transmission angle $\alpha_{\mathrm{tr}}$ notably varies with frequency $\Omega$ at the edges of Bragg reflection band (Fig. 2(b)) that is typical for linear reflection from periodic structures such as layered dielectric mirrors. In a high-frequency limit $\Omega \to \infty$ the transmission angle is almost independent on $\Omega$ in accordance with results of effective index approximation and approaches the value $\alpha_{\mathrm{tr}} = (\alpha_{\mathrm{in}}^2 + 2p)^{1/2}$. Notice that in deeper lattices the Bragg reflection band shifts into the high-frequency region (this is also consistent with the results of the analytical effective index approximation that predicts increase of $\alpha_{\mathrm{tr}}$, hence $\Omega_{\mathrm{B}}$, with $p$), while the maximal amount of reflected energy increases significantly. The shift of low-frequency edge of Bragg reflection band toward higher values of $\Omega$ with growth of the lattice modulation depth $p$ introduces an unexpected dependence of the reflected energy $U_{\mathrm{ref}}$ on $p$ for fixed $\alpha_{\mathrm{in}}$ and $\Omega$ (Fig. 2(c)). In this case the initial growth of reflected energy $U_{\mathrm{ref}}$ with increase of $p$ is replaced by rapid diminishing when the edge of the reflectance band shifts to the high-frequency region. This characteristic feature of Bragg-type soliton mirror is important, as the modulation depth as well as lattice frequency could serve as effective control parameters for routing. Figure 2(d) shows the dependence of the transmission angle on lattice depth. The dependence $\alpha_{\mathrm{tr}}(p)$ is nonmonotonic; rapid variations of $\alpha_{\mathrm{tr}}$ are obtained at the edges of Bragg-reflection band, illustrating the possibility to control not only the transmission coefficient, but also soliton propagation paths. The analytical effective index approximation (Fig. 2(d)) holds only in shallow lattices $(p < 1)$, stressing the differences existing between mechanisms of refraction at the interface of uniform media and interface with the lattice at $p > 1$. Figure 2(e) illustrates the dependence of transmitted energy (that was also calculated at the distance $\xi = 2\eta_0/\alpha_{\mathrm{in}}$) on the incident angle $\alpha_{\mathrm{in}}$. It is clear that the Bragg-type soliton mirror becomes transparent for sufficiently large incident angles. We also plotted in Fig. 2(e) the corresponding dependence $U_{\mathrm{tr}}(\alpha_{\mathrm{in}})$ calculated for the equivalent smoothed refractive index profile $R(\eta) = [1 + \tanh(\Omega\eta)]/2$ to stress the essential difference with transmission at the interface of two uniform nonlinear media. Finally, Fig. 2(f) shows the ratio of transmission and incident angles as a function of the incident angle. While at large $\alpha_{\mathrm{in}}$ one has $\alpha_{\mathrm{tr}} \approx \alpha_{\mathrm{in}}$, strong refraction is found at small incident angles, when soliton is launched almost parallel to the interface and a significant fraction of the transmitted energy is trapped in the near-surface guiding lattice channels. Importantly, this behavior is also in clear contrast with refraction at the interface between uniform media.

Therefore, in summary, we uncovered new important properties of soliton reflection and refraction at the interface formed by uniform and periodic nonlinear materials with focusing Kerr-type nonlinearity. Such Bragg-type nonlinear mirror offers new opportunities for control of reflection efficiency and beam propagation paths, because the refraction process is strongly sensitive to variations in depth and frequency of the periodic refractive index modulation. We stress the robustness of the result reported, in terms of particular values of physical parameters describing the interface, and the high soliton content of the reflected or transmitted beams.

**Acknowledgements**
This work was supported by Ramon-y-Cajal program and by CONACyT under project 46552.